\documentclass[%
 reprint,
 amsmath,
 amssymb,
 aps,
 pra,
]{revtex4-2}
\usepackage[utf8]{inputenc} 
\usepackage[T1]{fontenc}    
\usepackage{hyperref}       
\usepackage{url}            
\usepackage{booktabs}       
\usepackage{amsfonts}       
\usepackage{amssymb}        
\usepackage{nicefrac}       
\usepackage{microtype}      
\usepackage{lipsum}         
\usepackage{graphicx}       
\usepackage{xcolor}         
\usepackage{physics}        
\usepackage[detect-all]{siunitx} 
\usepackage{amsmath}
\usepackage{bm}             
\usepackage{cleveref}       
\usepackage{dcolumn}        
\usepackage{grffile}
\usepackage{wasysym}
\usepackage{import}         

\begin{document}

\title{Bacteriophages spreading through growing populations of motile bacteria: a theoretical and computational investigation}

\author{Rory Claydon}%
\author{Samuel Gartenstein}%
\author{Aidan T. Brown}%
\email{Corresponding author: aidan.brown@ed.ac.uk}
\affiliation{%
 School of Physics and Astronomy\\
  University of Edinburgh\\
  Edinburgh, Scotland\\
}
\date{\today}

\begin{abstract}

Bacteriophages spreading through populations of bacteria offer relatively simple, tuneable systems for testing mathematical models of range expansion. However, such models typically assume a static state into which to expand, which is not generally valid for bacterial-bacteriophage populations, where both the host (bacteria) and the infectious agent (bacteriophage) have similar growth rates. Here, we build on the classical FKPP theory of expanding fronts to study an infectious bacteriophage front propagating into an exponentially growing population of bacteria, focusing on the situation where the hosts are also mobile, e.g., swimming bacteria. In this case, both the infectious agent and the infected host populations take on the form of self-similar travelling waves with a fixed wave speed, as in FKPP theory, but the infected host wave also grows exponentially. Depending on the population under consideration, wave speeds are either advanced or retarded compared to the non-growing case. We identify a novel speed selection mechanism in which the shape of the bacteriophage wave controls these various wave speeds. We propose experiments to test our predictions.
\end{abstract}

\maketitle


\section{Introduction \label{Introduction}}

The growth of a population can have profound and unexpected impacts on processes within that population. For example, in bacterial colonies, population growth drives a rich variety of pattern formation mechanisms~\cite{bonachela2011, farrell2013mechanically}, leads to mechanical buckling~\cite{grant2014}, causes nematic defects within the colony to become self-propelled~\cite{dell2018growing, yamanEmergenceActiveNematics2019} and enables co-existence between the colony and bacterium-targeting viruses (bacteriophages, or `phages')~\cite{eriksen2018growing}. More generally, host population growth is predicted to reduce the basic reproduction number of an infection~\cite{may_endemicGrowing_1985}, while mutations are predicted to spread at higher speeds in populations that are themselves spreading in space~\cite{Rosalind_genetics_2014}. 

\begin{figure}[ht!]
\includegraphics[width=\linewidth]{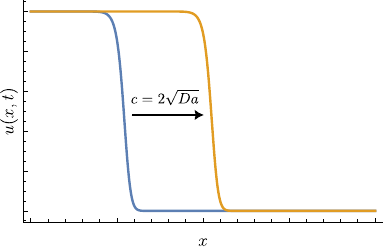}
\caption{
Numerical solution of \cref{eq:INTRO_Fisher-KPP} with $D=a=1$ and initial condition $u(x,0)=\exp(-t/2)/2$. A front propagating with $c=2\sqrt{Da}=2$ forms. 
\label{fig:FKPP_infection_wave}
}
\end{figure}

This last example illustrates the interaction between population growth and another ubiquitous phenomenon in ecology, the invasion of one unstable state by another more stable state~\cite{sarloos_frontPropRecent_2003, murray_mathematical_2003, metz_mollison_bosch_2000}. The paradigmatic model for such invasion problems is the Fisher-Kolmogorov-Petrovsky-Piskunov (FKPP) equation. In its original formulation~\cite{fisher_advanceGenes_1937, kolmogorov_KPP_1937} this reaction-diffusion equation described the fraction $u(x,t)$ of some advantageous mutant gene spreading through a population
\begin{equation}
\label{eq:INTRO_Fisher-KPP}
    \pdv{u}{t} = au\qty(1-u) + D\pdv[2]{u}{x}\,,
\end{equation}
with $u$ growing at rate $a$ towards a carrying capacity $u=1$ and diffusing at rate $D$. In the long-time limit and for most realistic initial conditions, the solution to \cref{eq:INTRO_Fisher-KPP} is a front translating uniformly at a characteristic speed, $2\sqrt{Da}$, which can be obtained by a linear expansion around the small mutant population at the extreme tip of the wave (see~\cref{fig:FKPP_infection_wave}). \Cref{eq:INTRO_Fisher-KPP} and its variants have found very wide application across numerous fields~\cite{sarloos_frontPropRecent_2003}, e.g., human history, particularly the spread of agricultural technology~\cite{ammerman_1971}, polymer physics~\cite{derrida_1988}, fluid dynamics~\cite{fineberg1987vortex}, computational search algorithms~\cite{majumdar_2002} and, as here, the spread of infectious diseases~\cite{abramson2003traveling, murray_mathematical_2003}.

\begin{table*}[ht!]
	\begin{tabular}{ c | c| c |c| c}
		Parameter & Symbol & Typical Range & Source & Default Value\\
		\hline
		Initial \emph{E.coli} conc. & $S_0$ & up to $\sim$\SI{1e9}{\per \milli\liter} & $^*$ & \SI{1e7}{\per \milli\liter} \\
		\emph{E. coli} growth rate & $a$ & up to $\sim$ \SI{1.7}{\per\hour} & \cite{scott2010}
		&\SI{1.9}{\per\hour}\\
		T4 lysis rate & $d$ $^\dagger$ &
		$\sim$0.5-\SI{2}{\per\hour} &
		\cite{vsivec_determinationBacteriophageGrwthParams_2020} & \SI{1.7}{\per\hour} $^\ddagger$\\
		Adsorption rate & $\beta$ & \SIrange[range-phrase = --,range-units = single]{4e-10}{4e-9}{\milli\liter\per\minute} & \cite{vsivec_determinationBacteriophageGrwthParams_2020}&\SI{4e-9}{\milli\liter\per\minute} $^\ddagger$\\
		Burst size & $n$  & up to $\sim$150 $^\S$ &
		\cite{vsivec_determinationBacteriophageGrwthParams_2020} & 4$ ^\parallel$\\
		\emph{E. coli} effective diffusivity & $D$ & \SIrange[range-phrase = --,range-units = single]{100}{130}{\micro\meter\tothe{2}\per\second} & \cite{schwarz-linek_2016} & \SI{130}{\micro\meter\tothe{2}\per\second}\\
		Phage diffusivity & $D_P$ & \SIrange[range-phrase = --,range-units = single]{3.4}{3.6}{\micro\meter\tothe{2}\per\second}& \cite{szermer_aggregationT4_2017,barr_subdiffusiveT4Diff_2015}& 0 $^\P$\\
		Superinfection parameter & $B_{\mathrm{sup}}$ & $0$ or $1$ & N/A & $1$ (superinfection on)\\
		Host-independent phage death rate & $m$ & $\sim{3\times 10^{-3}}\SI{}{\per\hour}$ & \cite{dePaepe2006} & $\SI{0}{\per\hour}$\\
		\hline
	\end{tabular}
	\caption{\label{tbl:parameters}
		Parameter values used in this paper for T4 infecting \textit{E. coli}. Where values other than the default are used this is indicated in the text. Notes: * \SI{2e9}{\per\ml} is a typical concentration above which \textit{E. coli} stops growing exponentially due to nutrient limitation, though this depends on the growth conditions and history. \SI{1e7}{\per\ml} is a typical concentration used to study exponential growth because it is sufficiently below the nutrient-limited concentration. $\dagger$ $d=1/L$ is obtained from the measured lysis period $L$. $\ddagger$ $\beta$ and $d$ both saturate above $a\sim$\SI{1}{\per\hour} so we take their maximum values. $\S$ Ref.~\cite{vsivec_determinationBacteriophageGrwthParams_2020} records a linear dependence between $a$ and $n$. The maximum value is obtained by extrapolating this dependence to $a=\SI{1.9}{\per\hour}$. $\parallel$~The significantly reduced burst size $n=4$ is to help accentuate the features discussed in \cref{Simplified model}; with the more realistic version of the model in \cref{realistic} we use the top of the range, $n=150$.  $\P$ Set to zero for simplicity because $D_P/D \ll 1$. We explore non-zero values of $D_P$ in \cref{sec:APP_extended_simple_model}.
}
\end{table*}

Analysis of FKPP-like equations relies on perturbation around the fixed, unstable initial state into which the front propagates \cite{sarloos_frontPropRecent_2003}. It is an open theoretical question whether the same phenomenology applies and what quantitative changes are necessary in the absence of this fixed initial state, e.g., for an infection spreading into a population which itself is growing. This question also has practical relevance. In human diseases there is often a separation of time scales between the host and viral reproduction rate so that the total population can be assumed constant~\cite{murray_mathematical_2003}, but this is not always the case, e.g., for chronic diseases like HIV or for countries with high birth rates~\cite{may_endemicGrowing_1985}. Similarly, no separation of time scales applies to the inter-microbe interactions that play an essential role in our global biochemical and geochemical cycles~\cite{danovaro2011}, e.g., bacteria and the viruses that infect them (bacteriophages, also known as phages) typically have similar growth rates~\cite{vsivec_determinationBacteriophageGrwthParams_2020}.

Extensive experimental and theoretical work has been conducted into the spread of bacteriophage infections through bacterial populations, typically focussed on phage `plaques,' i.e., the clearings formed by bacteriophages in bacterial lawns on semi-solid media such as agar~\cite{yin_replicationVirusPlaque_1992, kropinski_2009, fort2002time, Hunter_2021, Ping_2020}. These studies have produced quantitative predictions of wave speed~\cite{yin_replicationVirusPlaque_1992}, and have highlighted the impact on the infection dynamics of effects such as the distribution of phage lysis times~\cite{fort2002time} or bacterial crowding~\cite{Hunter_2021}. However, these studies have typically not been concerned with the impact of bacterial growth on the infection dynamics
~\cite{krysiak-baltynComputationalModelsPopulations2016}. 
In addition, apart from in a few cases, e.g, ref.~\cite{Ping_2020}, modelling has focused on the case where the bacteria are trapped within relatively hard agar, so that the mobility is provided solely by Brownian diffusion of bacteriophages.

Here, we will study the impact of exponential bacterial growth on the spread of bacteriophage infections. We will focus on the asymptotic wave speed of the infection and, as in ref.~\cite{Ping_2020}, allow for bacterial and bacteriophage mobility. In this paper, we want to stress the more mathematical and general aspects of this theory, so in~\cref{model}-\cref{analytical} we will keep the model as simple as possible, suppressing certain aspects of bacterial and bacteriophage behaviour. Nevertheless, we hope that our results will inspire experimental investigations into the impact of growth on infection speeds, so in~\cref{realistic}-\cref{sec: numerical} we will consider more general formulations of our model, which take into account features such as realistic distributions of the bacteriophage lysis time. We will also suggest a concrete experimental realisation to test our model, consisting of bacteriophages spreading through a thin, fluid-filled channel containing a population of growing bacteria.

Our main result is that the infected and uninfected bacteria form self-similar travelling waves, which retreat before the expanding phage front and which grow exponentially in time. The phage also form a self-similar front, which does not grow exponentially, but this is only in the case where superinfection (where a single bacterium can be simultaneously infected by multiple phage) is permitted; without superinfection the phage wave also grows and changes shape as it develops. The speeds of these various waves depend on the species tracked (bacteria or phage) and on whether the front or peak of the wave is tracked: the viral wave is retarded, while the wavefront of infected bacteria is advanced, compared to the case without bacterial growth. The advanced speed of the infected bacterial wave does not stem from the initial conditions, as is usual in FKPP theory, but is instead controlled dynamically by the shape of the phage wavefront in a novel selection mechanism. Interestingly, the varying wave speed also causes a non-monotonic variation in the width of the infectious wave, which is narrowest at intermediate growth rates.


\section{A General Theoretical Model for a Phage Infection Spreading Through an Exponentially Growing Host Population \label{model}}

The experimental situation we have in mind is a population of bacteriophages infecting a swimming bacteria such as \emph{Escherichia coli} in a quasi-1D system such as a thin, fluid-filled capillary. This would allow for bacteriophage and bacterial motion in 3D, but the motion of the wave would be restricted to 1D. We therefore model the dynamics of the number densities of susceptible bacteria $S$, infected bacteria $Q$ and phage $P$ in time $T$ and one spatial dimension $X$. Susceptible bacteria grow exponentially at rate $a$ and are infected by phage with rate constant $\beta$. Infected bacteria do not divide, but lyse after a time $\tau$ governed by a probability distribution $\Lambda(\tau)$, releasing $n$ phage when they do so. Bacteria typically have many phage-binding sites, so phage may in general superinfect bacteria, i.e., infect already-infected cells, also at rate $\beta$. However, some types of bacteriophage physically block the binding of a second bacteriophage to the cell~\cite{dulbeccoExclusion1952}
, so we will allow for this possibility in our model too. 

\textit{E. coli} cells swim with a `run-and-tumble' motion \cite{berg_rotary_2003,berg_chemotaxis_1972}, with straight-line motion (runs) punctuated by random changes of direction (tumbles). This gives long-time diffusive dynamics, with diffusivity $D \approx \flatfrac{v^2 t_r}{3}\sim\SI{130}{\micro\meter^2\second^{-1}}
$~\cite{lovely_1975}, where $v\sim \SI{20}{\micro\meter\second^{-1}}$ and $t_r\sim \SI{1}{\second}$ are the typical swimming speed and run duration, respectively~\cite{schwarz-linek_2016}. In principle, phage could modify the swimming speed of the infected bacteria, which we allow for by defining distinct diffusivities $D_S$ and $D_Q$ for the susceptible and infected bacteria, respectively. The phage diffuse due to Brownian motion, with a much lower rate $D_P\sim \SIrange{1}{10}{\micro\meter^2\second^{-1}}\ll D$  \cite{barr_subdiffusiveT4Diff_2015,szermer_aggregationT4_2017}.\\

We can represent this population dynamics through a set of coupled integro-differential equations
\begin{widetext}
\begin{subequations}
\label{subeq:MODEL_dimensional_system}
\begin{align}
\pdv{S}{T} &=\underbrace{a S}_{\text{growth}} - \underbrace{\beta PS}_{\text{infection}} + \underbrace{D_S \pdv[2]{S}{X}}_{\text{swimming}}\,,\\
\begin{split}
\pdv{P}{T} &= - \beta P (\underbrace{S}_{\text{infection}}+\underbrace{B_{\mathrm{sup}}Q}_{\text{superinfection}}) +\underbrace{D_P\pdv[2]{P}{X}}_{\text{diffusion}}-\underbrace{mP}_{\text{death}}\\
&\qquad + \underbrace{n\beta\int_{0}^T \dd{T'}\Lambda(T-T') \int_\infty^\infty \dd{X'} G_Q(X-X',T-T') P(X',T')S(X',T')}_{\text{lysis}}\,,
\label{eq:MODEL_P_full}
\end{split}
\\
Q(X,T) &=  \beta \int_{0}^T \dd{T'}  \qty(1-\int_0^{T-T'}\dd{T''}\Lambda(T''))\int_\infty^\infty \dd{X'} G_Q(X-X',T-T')
    P(X',T')S(X',T')\label{eq:MODEL_Q_full} \,,
\end{align}
\end{subequations}
\end{widetext}
where the superinfection parameter $B_{\mathrm{sup}}=1$ or $0$ with and without superinfection, respectively, and where the Green's function $G_Q(x,\tau)$ for the diffusion of infected bacteria is
\begin{equation}
\label{eq:infectives_diffusion_kernel}
 G_Q(x,\tau)=\left(4\pi D_{{Q}}\tau\right)^{-1/2}\exp(\dfrac{-x^2}{4D_{{Q}}\tau})\,.
\end{equation}
The meaning of the integral in \cref{eq:MODEL_P_full} is that the number of phage released by lysis at $(X, T)$ is an integral over all phage infection events at earlier times, weighted by the probability $G_Q(X-X',T-T')$ that the infected bacterium will reach position $X$ at time $T$ by diffusion and by the probability, $\Lambda(T-T')$, that the infected bacterium will lyse at time $T$ to release $n$ phage. Similarly, the integral in \cref{eq:MODEL_Q_full} means that the density of infected bacteria $Q(X,T)$ is given by the sum of all the prior infection events at earlier times $T'$ that arrive at $(X,T)$ by diffusion, and which have not already lysed. Note that $Q$ is an explicit function of $S$ and $P$, so that it can always be eliminated from the system of equations. This model is developed from similar integro-differential phage models~\cite{jonesBacteriophageBacteriaFlow2011, gourley_marineFoerster_2004} by allowing for unbounded exponential growth of the host population and a general lysis time distribution. In \cref{sec:APP_exponential_to_PDEs}, we derive \cref{eq:MODEL_Q_full} from first principles using the method in ref.~\cite{gourley_marineFoerster_2004} and show how this leads to the lysis term in \cref{eq:MODEL_P_full}.

\section{A simplified, partial-differential-equation model \label{Simplified model}}

We first consider a particular version of \cref{subeq:MODEL_dimensional_system} that can be written in the form of a set of coupled partial differential equations. This requires that the lysis time distribution is exponential, i.e., we write $\Lambda(T;d)=d\exp(-dT)$ with $d$ a fixed lysis rate. We also allow superinfection, so $B_{\mathrm{sup}}=0$ and we assume that $D_Q=D_S$ and that phage diffusion is negligible, so $D_P=0$. Then our model reduces to
\begin{subequations}
\label{subeq:MODEL}
\begin{gather}
\underbrace{\pdv{S}{T}}_{\mathrm{Susceptible}} = {\underbrace{a S}_{\mathrm{Growth}}} - {\underbrace{\beta P S}_{\mathrm{Infection}}} + {\underbrace{D_S\pdv[2]{S}{X}}_{\mathrm{Swimming}}}\,,\\
{
\underbrace{\pdv{Q}{T}}_{\mathrm{Infected}} = {\underbrace{\beta PS}_{\mathrm{Infection}}} - {\underbrace{d Q}_{\mathrm{Lysis}}} + {\underbrace{D_Q\pdv[2]{Q}{X}}_{\mathrm{Swimming}}}\label{subeq:MODEL_infected}}\,,\\
\underbrace{\pdv{P}{T}}_{\mathrm{Phage}} = {\underbrace{n d Q}_{\mathrm{Lysis}}} - {\underbrace{\beta PS}_{\mathrm{Infection}}} - {\underbrace{\beta PQ}_{\mathrm{Superinfection}}}\,,
\end{gather}
\end{subequations}
which is derived in \cref{sec:APP_exponential_to_PDEs}. We non-dimensionalize \cref{subeq:MODEL} by introducing dimensionless quantities: for the population densities, $s=S/S_0$, $q = Q/S_0$ and $p =P/S_0$, where $S_0$ is the initial, uniform bacterial population; for time, $t=\beta S_0 T$ and position, $x = X \sqrt{\beta S_0/D}$; and for the bacterial growth rate $\mu_s = a/(\beta S_0)$ and lysis rate $\gamma = d/(\beta S_0)$. The non-dimensional equations are then
\begin{subequations}
\label{subeq:MODEL_nondim}
\begin{gather}
\underbrace{\pdv{s}{t}}_{\mathrm{Susceptible}} = {\underbrace{\mu_s s}_{\mathrm{Growth}}} - {\underbrace{ps}_{\mathrm{Infection}}} + {\underbrace{\pdv[2]{s}{x}}_{\mathrm{Swimming}}}\,,\\
{
\underbrace{\pdv{q}{t}}_{\mathrm{Infected}} = {\underbrace{ps}_{\mathrm{Infection}}} - {\underbrace{\gamma q}_{\mathrm{Lysis}}} + {\underbrace{\pdv[2]{q}{x}}_{\mathrm{Swimming}}}\label{subeq:MODEL_infected}}\,,\\
\underbrace{\pdv{p}{t}}_{\mathrm{Phage}} = {\underbrace{n \gamma q}_{\mathrm{Lysis}}} - {\underbrace{p s}_{\mathrm{Infection}}} - {\underbrace{p q}_{\mathrm{Superinfection}}}\,.\label{subeq:MODEL_phage}
\end{gather}
\end{subequations}
To compare viral and host growth we define an effective (dimensionless) viral reproduction rate $\mu_p=\gamma (n-1)$, which is the rate at which a single viral particle would replicate in a large, uniform population of non-growing hosts, and $\mu=\mu_s/\mu_p$, the ratio of host to viral reproduction rates, which will emerge as our key control parameter.

\begin{figure}[ht!]
	\centering
	\includegraphics[width=\linewidth]{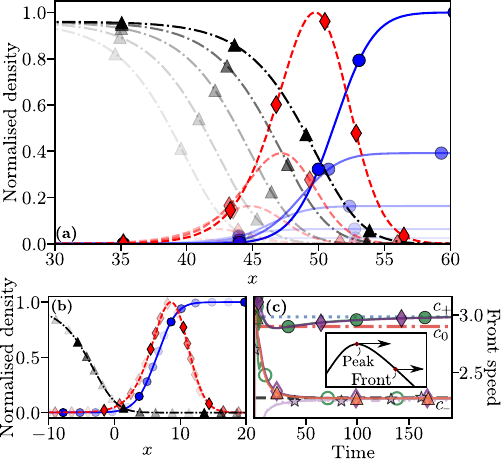}
	\caption{(a) 
Numerical solution to \cref{subeq:MODEL_nondim} at $\mu=0.8,~\gamma=0.7,~n=4$. All number densities are normalised by their maximum value at the final timepoint and earlier waves are more transparent, with symbols: $s~{\color{blue}{--}};~q ~{\color{red}{--}};~p~-$. $s$ and $q$ are self-simlar, exponentially growing wave whereas $p$ is a uniformly translating wave.  All population profiles are right travelling. (b) Data from (a) shifted in $x$ and re-scaled to show self-similarity with: $\sigma=se^{-\mu_s t}~{\color{blue}{--}};~\theta=qe^{-\mu_s t}~{\color{red}{--}};~p~-$. (c) Time dependence of wave speeds from (a). Front speeds: $p~ {\color[HTML]{f46d43}\blacktriangle};~\sigma~{\color[HTML]{1b7837}\Circle};~\theta~{\color[HTML]{762a83}{\lozenge}};~s~{\color[HTML]{1b7837}\CIRCLE};~q~{\color[HTML]{762a83}{\blacklozenge}}$; peak speed $q$ ${\color[HTML]{c2a5cf}{\bigstar}}$. Lines labelled $c_-~\text{--},~ c_0: {\color[HTML]{d73027}-\,{\scriptstyle\blacksquare}} ;~c_+~ {\color[HTML]{4575b4}{\scriptstyle\blacksquare}\,{\scriptstyle \blacksquare}}$ are as described in text. Inset: Schematic to distinguish peak and front speeds with logarithmic y axis 
\label{fig:1}}
\end{figure}

\section{Numerical and analytical results for the simplified model \label{analytical}}

Before analysing our model in detail, we will summarize the well understood theoretical features of the original FKPP equation, \cref{eq:INTRO_Fisher-KPP}. As described above, this equation supports self-similar wavelike solutions $u(\xi)$, with wave variable $\xi=x-ct$, that travel through the system at speed $c$ transforming the unstable initial state $u=0$ to the stable final state $u=1$. The front of the wave exhibits an exponential decay in space, $\lim_{\xi\rightarrow\infty}u\propto e^{-\lambda\xi}$, and the wave speed is coupled to the front steepness $\lambda$ through a dispersion relation $c(\lambda)$ that can be obtained via linear expansion around the initial state. There is a critical $\lambda=\lambda^*$ for which $c^*=c(\lambda^*)=2\sqrt{Da}$ is the minimal speed and only `shallow' waves with $\lambda\leq\lambda^*$ are stable: for initial conditions decaying more slowly than $e^{-\lambda^* x}$ the wave front matches this original front shape and travels at speed $c(\lambda)>c^*$, whereas for steeper initial conditions the original front decays into a critical wave of steepness $\lambda^*$ travelling at speed $c^*$. In practice, it can be shown that the discrete~\cite{derrida_1988} and spatially bounded~\cite{fisher_advanceGenes_1937} nature of populations mean that this minimal wave speed will almost always be selected, while stochastic effects lower the wave speed still further ~\cite{derrida_stochasticFrontProp_2001}. Extensions of~\cref{eq:INTRO_Fisher-KPP} to higher-order, difference, delay or integro-differential equations, or to multiple species, tend to yield similar phenomenology; these and numerous other theoretical results on the FKPP equation are reviewed in ref.~\cite{sarloos_frontPropRecent_2003}. 

\begin{figure}[hb!]
	\centering
	\includegraphics[width=\linewidth]{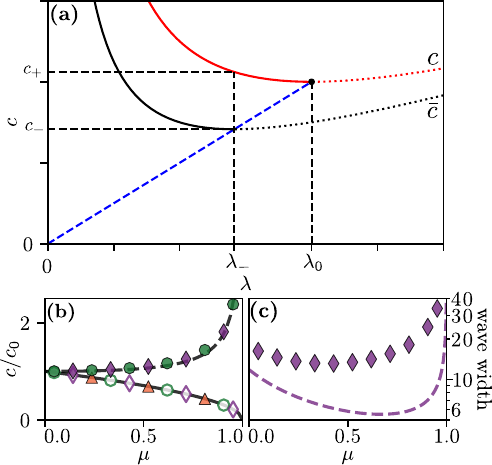}
	\caption{(a) Dispersion relations for: $q$ (red,~\cref{dispersion}); and $p$ (black~\cref{eq: dispersion c_-}) with $\mu=0.5$ and stable (solid) and unstable (dotted) regions indicated. With varying $\mu$ the minimum of $\bar{c}$ moves along the blue dashed line $c=2\lambda$ towards the origin. Other lines indicate speed selection mechanism as described in text. (b) Comparison between numerical (markers as in \cref{fig:1}c) and analytic (lines, $c_-$ solid, $c_+$ dashed) front speeds. Parameters are $n=4, \gamma=0.7$ and varying $\mu_s$. (c) The evolution of the infected wave width as a function of $\mu$ with the numerical FWHM (symbols) compared to the analytic approximation in~\cref{eq:approximate_predicted_width} (dashed curve).\label{fig:2}}
\end{figure}

Without bacterial growth,~\cref{subeq:MODEL_nondim} would be a generalized multiple-component FKPP equation~\cite{sarloos_frontPropRecent_2003} supporting travelling waves, with standard analysis predicting a wave speed $c_0=2\sqrt{\mu_p}$ in the high $S_0$ limit (see \cref{sec:APP_no_growth_wave_speed} for a derivation). With bacterial growth there is no unstable fixed point for waves to propagate into, so this standard FKPP analysis fails. We therefore perform numerical analysis of this system of equations. \Cref{fig:1}a shows a typical numerical solution of \cref{subeq:MODEL_nondim} obtained in Python using the backwards differentiation method~\cite{shampine_matlab_1997}. Parameter values, listed in \cref{tbl:parameters}, are here chosen to reflect the well studied T4 bacteriophages infecting \textit{E. coli}, but with a burst size of $n=4$, which is smaller than the typical T4 burst size of $\sim 100$ in optimal laboratory growth conditions~\cite{choi2010effect,RabinovitchModelBacteriophageT41999}. We choose this small burst size to better highlight the qualitative features of our mathematical model. In addition, the smaller burst size used here is realistic for T4 infecting slow-growing \emph{E. coli}~\cite{bryanBacteriophageT4Infection2016,RabinovitchModelBacteriophageT41999}, which are likely to be more representative of the near-starvation conditions of most bacteria in their natural environment \cite{hadasBacteriophagePhysiology1997,rabinovitchBacteriophageGrowth2002}. We will consider larger burst sizes in \cref{realistic}. 

The initial conditions were $[s,q]=[1,0]$ everywhere on the domain $x\in [0,~l_{sim}]$, with a smooth step-like bacteriophage profile near the origin $p(t=0)=p_0/(1+e^{\Upsilon(x-x_0)})$, with amplitude $p_0=5$, steepness $\Upsilon=10$, width $x_0=l_{sim}/5$ and no-flux boundary conditions. For full simulation details see \cref{sec:APP_SIM_DETAILS}. 
Propagation speeds, obtained numerically in the front (exponentially decaying) regions for each wave and at the peak for the infected wave are shown in \cref{fig:1}. The front speeds of $q$ and $p$ are obtained from the region where the height of each population is $10^{-6}$ of its peak height, and for the susceptibles we apply the same approach to the inverted population $\phi=e^{\mu_s t}-s$, see \cref{sec:APP_SIM_DETAILS} for details. 

We see in~\cref{fig:1}a that the phage exhibit a self-similar travelling wave profile, whereas the bacterial populations are self-similar but grow exponentially as $s,q\sim e^{\mu_s t}$. By scaling the bacterial populations as $\sigma=se^{-\mu_s t}$ and $\theta=qe^{-\mu_s t}$, we show $\sigma$, $\theta$, $p$ form a set of self-similar, uniformly translating waves in~\cref{fig:1}b. 
We investigate the front speeds of the original populations ($s$, $q$, $p$) and also of the re-scaled bacterial populations ($\sigma$, $\theta$) in \cref{fig:1}c. The measured speeds separate into two groups: $c_+$ ($s$ and $q$ front speeds); and $c_-$ ($p$ front and $q$ peak speeds). 


We therefore first analyze the problem in terms of the re-scaled populations. As is typical with population expansion problems, we will focus mainly on the long-time asymptotic behaviour, particularly the wave speed. As we will discuss in~\cref{sec: numerical}, we can expect to achieve reasonable convergence towards these long-time asymptotics even within experimental timescales, which are typically limited by the nutrients available to the bacteria. 


In this long-time limit, and at the front of the wave ($x\rightarrow\infty$), $s$ is very large, so the phage-binding term $ps$ dominates \cref{subeq:MODEL_phage}. Hence, $p$ relaxes more quickly towards equilibrium than $s$ or $q$, and we can therefore approximately replace $p$ by its steady-state value
\begin{align}
\label{eq:p_eqm}
p\sim n\gamma q/(s+q) = n\gamma \theta/(\sigma+\theta)\,.    
\end{align}
Interestingly, this approximation, as we verify numerically in \cref{sec:APP_phg_eqm}, applies even at the rear of the wave where $ps$ no longer dominates. This is because $q$ and $s$ are both small at the rear of the wave, so that $\partial p/\partial t$ vanishes, and the phage population retains the steady-state value it had attained at the front of the wave. Inserting~\cref{eq:p_eqm} into~\cref{subeq:MODEL_nondim}b and transforming $s,~q$ into $\sigma,~\theta$ gives
\begin{subequations}
\label{subeq:RandD_scaled_system2}
\begin{align}
\pdv{\sigma}{t} &= -\frac{n\gamma\theta\sigma}{\theta+\sigma} + \pdv[2]{\sigma}{x}\,,\\
\pdv{\theta}{t} &= \frac{n\gamma\theta\sigma}{\theta+\sigma}-\qty(\gamma+\mu_s) \theta + \pdv[2]{\theta}{x}\,,
\end{align}
\end{subequations}
which now does have an unstable fixed point at $[\sigma,~\theta]=[1,~0]$ (in fact, there is a continuum of fixed points along the line $\theta=0$, but this distinction is irrelevant here). Hence we expect self-similar waves, as seen in~\cref{fig:1}b, with a single speed determined by linearizing around the unstable fixed point. For the infected class this linearization gives
\begin{equation}
\label{linearized_theta}
\pdv{\theta}{t}=\qty[(n-1)\gamma-\mu_s] \theta + \pdv[2]{\theta}{x}\,,
\end{equation}
which is indeed the linearized form of the FKPP equation (c.f., \cref{eq:INTRO_Fisher-KPP} with $u\ll1$). Inserting the ansatz $\theta\sim e^{-\lambda\xi}$, yields the dispersion relation
\begin{align}
\label{eq: dispersion c_-}
\bar{c}(\lambda)=\lambda+\mu_p(1-\mu)\lambda^{-1}\,,
\end{align}
where $\bar{c}$ indicates the wave speed in re-scaled population space, i.e., $\sigma,~\theta,~p$ space. As is standard for FKPP analysis, we expect the system to choose the minimal speed and the maximum stable steepness because the initial conditions are sufficiently steep~\cite{sarloos_frontPropRecent_2003}. This gives
\begin{equation}
\label{eq:RandD_scaled_velocity}
    c_- =\min\bar{c}= 2 \sqrt{\gamma(n-1) - \mu_s}=c_0\sqrt{1-\mu}\,,
\end{equation}
and $\lambda_-=c_-/2$, see the lower curve in \cref{fig:2}a. We recall here that $\mu=\mu_s/\mu_p=\mu_s/[\gamma(n-1)]$ is the bacteria:phage growth-rate ratio. Because of self similarity, the front of $\sigma$ and the peak of $\theta$, which is also the peak of $q$, move at speed $c_-$ and have steepness $\lambda_-$.

Returning to the un-scaled populations, the higher front speed of the $s,~q$ waves is then explained by their exponential growth. We substitute back $q=e^{\mu_s t}\theta$ and define $q\sim e^{-\lambda_-\xi_+}$ at the front, with $\xi_+=x-c_+t$, which corresponds to a wave translating uniformly. This yields a speed $c_+\geq c_0$ given by
\begin{equation}
\label{eq:RandD_correct_speed_q_s}
    c_+=\frac{2\gamma(n-1)-\mu_s}{\sqrt{\gamma(n-1)-\mu_s}}=\frac{c_0\left(1-\mu/2\right)}{\sqrt{1-\mu}}\,.
\end{equation}
We note that this higher front speed is in some sense an artefact: it arises because we define the front speed in isolation, as if the wave behind were not growing. Nevertheless, it represents an experimentally relevant quantity, e.g., the largest $x$-position where we can detect a measurable concentration of infected bacteria will advance at speed $c_+$.

These predictions agree with the numerics in \cref{fig:1}, verified for $\mu<1$ in \cref{fig:2}b. For $\mu\geq1$ non-physical complex speeds are predicted. From~\cref{subeq:RandD_scaled_system2}b we see that the fixed point at $[\sigma,~\theta]=[1,~0]$ becomes stable for $\mu>1$ and we verify numerically that waves are not supported; instead the bacteria continue growing exponentially because the phage replicate too slowly to overtake them, see~\cref{sec:APP_mu_greater_1}. Hence there is a transition from a state that supports travelling waves to one that does not at $\mu=1$.

Alternatively, we can analyse the wave speeds entirely in the un-scaled population space, and this will reveal a novel speed selection mechanism. First, if we naively apply the approximation in~\cref{eq:p_eqm} to the original infected cell equation~\cref{subeq:MODEL_nondim}b in spite of the absence of a fixed point, and linearize by taking $s\gg q\sim 0$, this yields a FKPP-type equation for the infected cells
\begin{equation}
\label{infected cell_unscaled}
\pdv{q}{t} = (n-1)\gamma q+\pdv[2]{q}{x}\,,
\end{equation}
which yields the dispersion relation
\begin{align}
     \label{dispersion}
     c(\lambda)=\lambda+\mu_p\lambda^{-1}\,.
\end{align}
\Cref{dispersion} has the minimal point $[\lambda,~c]=[\lambda_0,~c_0]$, with $\lambda_0=c_0/2$, see~\cref{fig:2}a, upper curve. We might therefore expect the infected wave to travel at speed $c_0$, in contradiction to the numerics and our previous analysis. This failure is at first surprising, since \cref{infected cell_unscaled} has the standard linearized FKPP form. However, we must also examine the phage wave. If we apply the same analysis to the phage population by rearranging \cref{eq:p_eqm} to substitute for $q$ in \cref{subeq:MODEL_nondim}b we see that the phage dispersion relation is given by the lower curve $\bar{c}(\lambda)$ in~\cref{fig:2}a, i.e., by~\cref{eq: dispersion c_-}, which does have minimal speed $c_-$ and critical steepness $\lambda_-$, in agreement with our previous analysis. Then, if we take the form of the phage wave at the wavefront, $p\propto e^{-\lambda_- x}$, and insert this into~\cref{eq:p_eqm} we obtain $q\sim sp/(n\gamma)\propto e^{\mu_s t}e^{-\lambda_- x}$ for the infected population, exhibiting the same shallow decay, $\sim e^{-\lambda_-x}$, as the phage front. Hence, this imposes a steepness $\lambda_-$ on the infected wave, and calculating the wavespeed from~\cref{dispersion} gives $c(\lambda_-)=c_+$ in agreement with our previous calculations. This is represented graphically in~\cref{fig:2}a. 

In other words, the phage dispersion relation controls the overall dynamics of the system; this then imposes a shallow decay on the infected cell wave, generating a higher wave speed. Whereas in the standard FKPP analysis the wave is selected that has the maximum stable steepness, in our coupled system the selected wave has the maximum steepness that is stable for \emph{both} populations. This represents a novel speed-selection mechanism: the infected cell wave speed is not determined by the initial conditions but by a front shape that emerges from the dynamics of the system itself. This mechanism is reminiscent of that identified in ref.~\cite{Rosalind_genetics_2014}, in which a mutation spreads through a population which itself is spreading spatially. There, the exponential front of the entire population wave provided a slowly decaying front, which acts as an initial condition for the wave of genetic modification that followed, causing that second wave to accelerate. This is distinct from our mechanism: in ref.~\cite{Rosalind_genetics_2014} there is no back-coupling between the overall population wave and the mutation wave, so the overall population wave would travel at a fixed rate independent of the mutational dynamics behind it and even in the absence of mutations. In our system the two wave speeds $c_-$ and $c_+$ are instead generated by a two-way interaction between bacteria and phage.


We now briefly examine the shape of the infected wave. Measuring the wave's full width at half maximum (FWHM) numerically reveals a non-monotonic dependence on the relative growth rate $\mu$, see \cref{fig:2}c. This can be explained by examining the rear of the wave in re-scaled population space. All waves travel at speed $c_-$ in the re-scaled space, so the populations are functions of the wave variable $\xi_-=x-c_-t$. In the rear of the wave there are many more infected than susceptible cells, so $\theta\gg\sigma$ for $x\rightarrow-\infty$. Solving \cref{subeq:RandD_scaled_system2}b in this limit we obtain an exponential decay towards negative $x$-values: $\theta(\xi_-)\sim \exp( \chi \xi_- )$, with a new steepness parameter $\chi = \sqrt{ n \gamma } - \lambda_-\sim\lambda_0\left(1-\sqrt{1-\mu}\right)$, where the approximation is for $n\gg1$. An approximate FWHM, taking into account just the front and rear exponential regions is
\begin{equation}
\label{eq:approximate_predicted_width}
    w_{approx} 
    =\frac{\lambda_0^{-1}\ln 2}{\sqrt{1-\mu}\left(1-\sqrt{1-\mu}\right)} \,,
\end{equation}
which we plot in \cref{fig:2}c. This approximation captures the qualitative behaviour of the wave: as the wave speed decreases with increasing $\mu$ the wave is compressed at the front and expanded at the rear, which results in a minimal width at intermediate growth rates. The approximation significantly underestimates the numerical width by ignoring the peak region itself, but we could obtain arbitrarily good agreement by choosing a wider definition of the width, e.g., full width at tenth maximum, where the central region becomes negligible. 

\section{Modifications to the Simplified Model \label{realistic}}

In this section we consider some extensions and modifications to the simplified model presented above, focusing again on the asymptotic wave speed. In \cref{sec: numerical} we will present numerical results on a more general model that should better reflect the experimental situation of bacteriophages spreading through a population of growing, planktonic bacteria.

First, the theoretical limiting wave speeds predicted by the simplified model do not change if we allow any or all of: varying bacterial diffusivities, $D_S\neq D_Q$; a phage diffusivity, $D_P>0$; or a bacteria-independent phage death rate, $m>0$. In all cases, the wave speed is governed by the infective diffusivity $D_Q$, which replaces $D$ in the equations above. This can be shown by linearizing around the wave front, as before, see \cref{sec:APP_extended_simple_model}. The parameters apart from $D_Q$ drop out because of the high susceptible concentration at the front: the susceptibles act as a constant background, unaffected by the spreading wave, so there is no impact of $D_S$; likewise, new phage bind almost instantly to a new host so there is no time for phage diffusivity or bacteria-independent decay to have an impact. Notably, when $D_Q,~D_S=0$ we predict that growth will produce a vanishing wave speed even for $D_P>0$. This is relevant for models of the standard plaque-assay phage-counting technique, which relies on diffusion of phages through a population of immobilized bacteria~\cite{kropinski_2009, fort2002time, yin_replicationVirusPlaque_1992}. Some of these models also predict a vanishing wave speed in this high-bacterial-concentration limit~\cite{yin_replicationVirusPlaque_1992}. The qualitative explanation is that at high bacterial concentrations phage spend all their time bound to static bacteria and have no time to diffuse, a point also noted in ref. \cite{gourley_marineFoerster_2004}.  

Second, if we repeat our theoretical calculations with a general lysis time distribution $\Lambda(T)$, our predictions remain unchanged except that the key parameter $\mu=\mu_s/\mu_p$ is now determined by an integral over the lysis time distribution, see \cref{app: front speed general model} for details. In particular, using a more realistic delta-function distribution $\Lambda(T)=\delta(T-L)$ with $L$ a fixed lysis time, as in ref.~\cite{gourley_marineFoerster_2004}, gives $\mu=a L/\ln(n)$. For realistic parameters (\cref{tbl:parameters}), now including $n=150$, rather than $n=4$, we obtain a dimensional wave speed $c_0\sqrt{\beta S_0 D}=2\sqrt{\ln(n)D/L}\approx\SI{1}{\micro\meter\second^{-1}}$ for relative growth rate $\mu\sim 0.2$, and a speed difference $(c_+-c_-)/c_-~\sim 9\%$, which should be easily observable experimentally. We note that, for the same parameters in the minimal model of~\cref{Simplified model}, we would obtain a much smaller speed difference $\sim 1\%$ because $\mu\propto n^{-1}$: hence we used a lower $n=4$ throughout that section to illustrate the wave-speed difference graphically. 

We can simulate the delta-function distribution easily using a delay-differential-equation (DDE). However, this gives large oscillations in the wavespeed, see \cref{sec:APP_PIDE_no_phage_diffusion}. These oscillations are maintained for the entire duration of our simulations, which is probably because the phage released at a given time immediately infect new bacteria and therefore do not interact with other bacteria or phage until they are released after the fixed delay time. Hence, the phage released at any given time $T$ are effectively uncoupled from phage released at later times, apart from periodically at $T+L$, $T+2L$ etc, which means there is no damping term. The instability underlying these oscillations is theoretically interesting and worthy of further study. However, they are unlikely to be visible in real biological experiments where natural heterogeneity in all parameters will presumably dampen them. As a final note, periodic solutions were observed in a similar system \cite{jonesBacteriophageBacteriaFlow2011} and in general,  oscillatory solutions are expected for DDEs e.g. \cite{murrayMathematicalBiology2002}.

Third, if we remove the superinfection term from \cref{subeq:MODEL_nondim}, numerical solutions show the phage now grow unboundedly, see \cref{fig:no_superinfection}, as there is no longer any mechanism by which phage may be removed from circulation, other than by infecting a susceptible bacterium. As this only affects the bulk of the wave behind the front, there is no effect on the front speed and the asymptotic wave speed is still realised. The phage wave as a whole is no longer self-similar but the bacterial dynamics remain unaffected. 

Fourth, the continuum assumption will break down near the wave front where the population is low. For the standard FKPP equation the resulting stochasticity introduces a speed reduction $\propto \ln^{-2}(N)$, with $N$ the approximate number of particles in the wave front~\cite{derrida_stochasticFrontProp_2001}. In our model the bacterial population grows exponentially so one might expect this correction to vanish with time. However, the phage wave retains a constant population at the front, and as it is this wave which determines both wave speeds through the steepness $\lambda_-$, it seems probable that some stochastic correction will remain.

\section{Numerical calculations for a more realistic model \label{sec: numerical}}

Our model cannot be tested in the usual environment of an agar plate: for such systems, where the phage are mobile rather than the bacteria, we predict that the asymptotic wave speed in populations of growing bacteria will vanish. A more suitable experimental setup for testing our theoretical predictions would therefore be a fluid-filled channel containing a suspension of swimming bacteria into which phage are inserted at one end. The various wave speeds and shapes would then be accessible via microscopy or light scattering. Experimental parameters could be controlled, e.g., through the nutritional quality ~\cite{vsivec_determinationBacteriophageGrwthParams_2020} or viscosity~\cite{martinez_2014} of the medium. Since we have mainly been interested in the long-time asymptotics, a natural concern is whether experiments will approach the asymptotic behaviour before the bacteria run out of nutrients. This section will answer this question through numerical calculations based on bacteriophage with a realistic lysis-time distribution, and other parameters chosen to match experimental data in the literature.

For the lysis time distribution we use the shifted Gamma distribution
\begin{equation}
\label{eq:shiftedGamma}
\Lambda(T)=
\begin{cases}
0  & T<T_e\,, \\
\frac{r^\alpha}{\Gamma(\alpha)}T^{\alpha-1}e^{-r (T-T_e)} & T\geq T_e\,. 
\end{cases}
\end{equation}
which replicates most of the essential features of real bacteriophage lysis-time distributions: there is a broad, continuous distribution of lysis times following a period, the so-called `eclipse' period, when there is strictly zero probability of lysis. If $\alpha$ is an integer, this also allows us to reproduce the distribution via a coupled system of PDEs which steps through a set of $\alpha+1$ infected classes at a fixed rate, $r$, as in ref.~\cite{Ping_2020}. This is simpler to implement than integro-differential equations. Here, we use $\alpha=2$. The lysis period, $L$, corresponds to the average time taken to progress through all stages of infection, and so $L=T_e+2/r$ ($L=T_e+\qty(\alpha-1)/r$ in general). We use $T_e=\SI{22.5}{\minute}$ and $1/r=\SI{5}{\minute}$ for biologically relevant parameters~\cite{RabinovitchModelBacteriophageT41999}. This gives $L=\SI{32.5}{\minute}$, which is a slight modification from the default value of $L\approx\SI{37}{\minute}$ in \cref{tbl:parameters}. Simulation details are given in \cref{sec:APP_realistic_model_sim_details}.

\begin{figure}
\centering
\includegraphics[width=\linewidth]{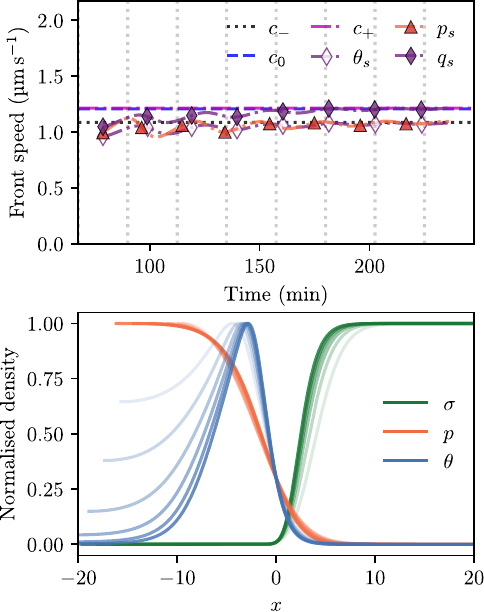}
\caption{
(a) Smoothed front speed data from numerical solutions of \cref{subeq:MODEL_dimensional_system} using the default parameters from \cref{tbl:parameters}. The speed data has been averaged with a sliding window over periods equal to the lysis period for clarity, with the subsript $s$ added to indicate smoothed data. Eclipse periods are delineated by the dotted vertical lines, 22.5 minutes apart for the default parameter set. The original wave speeds are plotted in \cref{fig:APP_PIDE_full_wave_speeds}.
(b) Example waves from which speed data in (a) was extracted. These waves correspond to the last two eclipse periods in (a). To show these converge to self similar waves, they have been normalised by their maximum and then aligned by the same height on the front. The more faded the wave, the further in the past it is.
\label{fig:DDE_self_similar_speeds}}
\end{figure}

\Cref{fig:DDE_self_similar_speeds}a mirrors \cref{fig:1}a, showing that the infected and susceptible population waves approach self-similar travelling waves after an initial transient period. The front portion of all population waves always approaches a self-similar structure \footnote{Depending on parameters, the phage wave may not be self-similar in the rear of the wave, far from the front. These perturbations do not propogate and the front of the wave converges to a self-similar profile.}. In \cref{fig:DDE_self_similar_speeds}b we extract the smoothed front speed, which agrees with the theoretical front propagation speeds $c_-$ and $c_+$ for this model, calculated in \cref{app: front speed general model}. Note that the wavespeeds converge to within a few percent of their theoretical values within 125 minutes, with full convergence by 180 minutes. For the default simulated bacterial growth rate and starting concentration from \cref{tbl:parameters}, a real bacterial suspension would reach stationary phase at around 170 minutes, assuming the typical maximum concentration of \SI{2e9}{\milli \litre^{-1}} at which \emph{E. coli} enter into the stationary phase~\cite{finkel2006long}. Hence, we can expect to observe both the development of self-similar waves and the splitting of the measured wave speeds within typical experimental timescales. Even longer observation periods could be achieved by performing experiments in thin channels that can exchange nutrients and waste with a reservoir, e.g., channels microfabricated in agarose~\cite{Priest_2017}.

\section{Discussion and conclusion}

In this paper we have made several theoretical predictions for the spread of a phage infection in an exponentially growing bacterial population: the existence of self-similar (phage) and exponentially growing (bacterial) travelling waves; the various speeds exhibited by those waves; and a non-monotonic relationship between the growth rate and the width of the infected cell wave. These predictions should be testable experimentally by allowing bacteriophage and bacteria to interact in a fluid-filled capillary, and we expect to test these results experimentally in a subsequent publication. 

We focused here on the asymptotic wave speeds obtainable theoretically. Throughout, these wave speeds matched the predictions of FKPP theory, implying that these are \emph{pulled} waves, i.e., driven by the infection dynamics in the very tip of the wave. This contrasts with recent work on bacteriophage plaques~\cite{Hunter_2021} where some conditions exhibited \emph{pushed} waves, which are faster waves driven by growth in the body of the wave. It will be interesting to explore whether this absence of pushed waves is a generic feature of the type of model studied here, where the virus spreads principally through bacterial motility. It would also be interesting to explore the impact of our results on genetic diversity. However, we might expect this effect to be small: in general, only the small population at the front of a pulled wave~\cite{giomettoEmergingPredictableFeatures2014,wakitaExperimentalInvestigationValidity1994,panjaEffectsFluctuationsPropagating2004} is able to contribute to genetic evolution, and the size of this front population will be governed by the decay length of the spreading wave, which itself is not predicted to be significantly affected by bacterial growth in realistic experimental conditions.

Finally, variations on this model will also likely be applicable to other systems where growth and invasion occur on similar time scales, e.g., chronic diseases, technological developments and mutations spreading through exponentially growing human or animal populations, or, as with the FKPP equation itself, in wholly unexpected fields. The benefit of the bacteria/bacteriophage system is that it is readily accessible to experiment.

\begin{acknowledgments}
RC acknowledges funding from the Scottish Universities Physics Alliance (SUPA) and the UK EPSRC through the Condensed Matter Centre for Doctoral Training (CM-CDT, EP/L015110/1). SG was supported by a University of Edinburgh Higgs Centre Prize MSc Scholarship. AB acknowledges funding from an EPSRC Innovation Fellowship (EP/S001255/1). We would like to thank M. Evans, R. Allen, F. Bull and J. de Graaf for useful discussions. For the purpose of open access, the author has applied a Creative Commons Attribution (CC BY) licence to any Author Accepted Manuscript version arising from this submission. 
\end{acknowledgments}

\onecolumngrid
\appendix
\counterwithin{figure}{section}
\counterwithin{table}{section}
\section{Derivation of the death rate of infectives} \label{sec:APP_exponential_to_PDEs}

In the main text we claim that \cref{subeq:MODEL} is a special case of \cref{subeq:MODEL_dimensional_system} if the lysis distribution is chosen to be $\Lambda(t;\gamma)=\gamma\exp(-\gamma t)$ with $\gamma$ a fixed lysis rate. We begin by deriving the loss term due to lysis of the infected bacteria in \cref{subeq:MODEL_dimensional_system} by introducing the age structure of the infected bacteria and solving the resulting von Foerster equation as in ref. \cite{gourley_marineFoerster_2004}. We show how this death term and the solution from the von Foerster equation transforms \cref{subeq:MODEL_dimensional_system} to the simplified model \cref{subeq:MODEL}.

The following derivation is a slight generalisation of the method in \cite{gourley_marineFoerster_2004} used to derive the lysis term of the infected bacteria using a Fourier transform approach. We generalise the derivation by allowing for a general hazard function of the infected bacteria, which represents the rate at which infected bacteria are lost as a function of age. Following the approach in ref. \cite{gourley_marineFoerster_2004}, we define the density of infectives at $(x,t)$ of age $a$ to be $\mathcal{Q}(x,t,a)$. The total number of infected bacteria at $(x,t)$ is then given by $Q=\int \dd{a} \mathcal{Q}(x,t,a)$. The evolution of the age structured model is governed by the Foerster equation
\begin{equation}
\label{eq:Foerster_full_pdv_Q}
    \pdv{\mathcal{Q}}{t} + \pdv{\mathcal{Q}}{a} = D_{\mathcal{Q}}\pdv[2]{\mathcal{Q}}{x} - m(a) \mathcal{Q}\,,
\end{equation}
where $D_{\mathcal{Q}}$ is the diffusion constant of the infectives and $m(a)$ the death function or hazard rate, which is the rate at which infected bacteria of age $a$ are lost. If lysis is the only mechanism by which infected bacteria may be lost, then $m(a)$ is related to the lysis time probability distribution, $\Lambda(a)$, through
\begin{equation}
    m(a) = \frac{\Lambda(a)}{1-\int_0^a \dd{a'} \Lambda(a')}\,,
\end{equation}
or more usefully
\begin{equation}
\label{eq:lysis_distribution_relation_to_hazard}
\Lambda(a)=-\pdv{a}\qty( \exp(-\int_0^{a}\dd{a'}m(a')))\,.
\end{equation}
The age $a$ is measured from the infection event, which gives the initial condition
\begin{equation}
    \mathcal{Q}(x,t,0) = \beta S(x,t) P(x,t)\,.\label{eq:app_birth_infectives}
\end{equation}
The standard solution procedure for solving Von Foerster equations is to use the method of characteristics \cite{vonFoerster_1959,murrayMathematicalBiology2002}. We reparameterise the system with $\alpha$, requiring $\dv*{t}{\alpha}=\dv*{a}{\alpha}=1$. The characteristic curves then have the relation $a=a_0+\alpha$, $t=t_0+\alpha$ and so taking the convenient choice $a_0=0$ allows us to use the initial condition $\mathcal{Q}(x,t,0)$ from above. Denoting $\mathcal{Q}_\alpha(x,\alpha)\equiv\mathcal{Q}(x,t(\alpha),a(\alpha))$ one finds
\begin{equation}
\label{eq:Foerster_barQ_pdv}
    \pdv{\mathcal{Q}_\alpha}{\alpha} = \pdv{\mathcal{Q}_\alpha}{t} + \pdv{\mathcal{Q}_\alpha}{a} = \qty( D_{\mathcal{Q}}\pdv[2]{}{x} - {m}(a)) \mathcal{Q}_\alpha\,.
\end{equation}
Continuing by taking the Fourier transform of \cref{eq:Foerster_barQ_pdv} in $x$, one obtains
\begin{equation}
    \pdv{\tilde{\mathcal{Q}}}{\alpha} = \qty(-D_{\mathcal{Q}} k^2 - m(\alpha)) \tilde{\mathcal{Q}}
\end{equation}
which has the solution
\begin{equation}
\begin{split}
\tilde{\mathcal{Q}}(k,\alpha) &= \tilde{\mathcal{Q}}(k,0) \exp(-D_{\mathcal{Q}} k^2 \alpha - \int_0^{\alpha} \dd{\alpha'}m(\alpha'))\\
&=\tilde{\mathcal{Q}}(k,0) \mathcal{G}(k,a)\,,
\end{split}
\end{equation}
where we have identified $\mathcal{G}(k,a)$ as the Fourier transform of the Green's function. We can find the Fourier transform of the initial conditions, \cref{eq:app_birth_infectives}, to give $\tilde{\mathcal{Q}}(k,0) = \mathcal{F}\qty[\beta S(x,t_0) P(x,t_0);x\rightarrow k]$.We proceed to take the inverse transform and, by the convolution theorem 
\begin{equation}
   \mathcal{Q}(x,t,a)=\mathcal{Q}_\alpha(x,\alpha)= \beta \int_{\infty}^{\infty} \dd{y} \mathcal{G}(x-y,\alpha)  S(y,t_0) P(y,t_0)\,.
\end{equation}
More usefully
\begin{equation}
   \mathcal{Q}(x,t,a)= \beta \int_{\infty}^{\infty} \dd{y} \mathcal{G}(x-y,a)  S(y,t-a) P(y,t-a)\,,
\end{equation}
and so 
\begin{align}
    Q(x,t) &= \int_0^{\infty} \dd{a}\mathcal{Q}(x,t,a)\\
           &=  \beta \int_0^{\infty} \dd{a} \int_{\infty}^{\infty} \dd{y} \mathcal{G}(x-y,a)  S(y,t-a) P(y,t-a)\\
           &=  \beta \int_{-\infty}^{t} \dd{\tau} \int_{\infty}^{\infty} \dd{y} \mathcal{G}(x-y,t-\tau)  S(y,\tau) P(y,\tau)\,.\label{eq:app_Q_explicit}
\end{align}
In summary, an explicit partial differential equation for the infectives is not required as it is entirely determined by the equations for $S$ and $P$.

Explicitly,
\begin{align}
    \mathcal{G}(x,t) &=  \frac{\exp(-\flatfrac{x^2}{4D_{\mathcal{Q}}t})}{\sqrt{4D_{\mathcal{Q}}\pi t}}
   \exp(-\int_0^{t}\dd{\alpha}m(\alpha))
   \,.
\end{align}
To determine the death rate through lysis for the infected bacteria, we need to find the evolution equation for $Q(x,t)$. This can be done by integrating \cref{eq:Foerster_full_pdv_Q} over $a$, and using \cref{eq:app_Q_explicit}. This gives
\begin{equation}
\begin{split}
 \pdv{Q}{t} &=  \pdv[2]{Q}{x}+\beta P(x,t)S(x,t) \\
    &- \beta\int_{0}^{\infty}\dd{a} \Lambda(a)\int_{-\infty}^{\infty}\dd{y}\mathcal{G}_D(x-y,a)P(y,t-a)S(y,t-a)\,,
    \end{split}\label{eq:app_death_lysis}
\end{equation}
where we made use of \cref{eq:app_birth_infectives,eq:lysis_distribution_relation_to_hazard} and $Q(x,t,\infty)=0$, i.e., infected bacteria do not live forever.

Taking an exponential distribution for lysis, e.g. 
\begin{equation}
\Lambda(a)=\gamma e^{-\gamma a}\,,\label{eq:app_exp_lysis}
\end{equation}
gives the lysis death rate from \cref{eq:app_death_lysis} as
\begin{equation}
\beta\gamma\int_{0}^{\infty}\dd{a} e^{-\gamma a}\int_{-\infty}^{\infty}\dd{y}\mathcal{G}_D(x-y,a)P(y,t-a)S(y,t-a)\,,
\end{equation}
which comparing to $Q(x,t)$ from \cref{eq:app_Q_explicit} with \cref{eq:app_exp_lysis} substituted,
\begin{equation}
Q(x,t) = \beta\int_{0}^{\infty}\dd{a} e^{-\gamma a}\int_{-\infty}^{\infty}\dd{y}\mathcal{G}_D(x-y,a)P(y,t-a)S(y,t-a)\,,
\end{equation}
shows that the evolution equation for $Q$ may be written
\begin{equation}
 \pdv{Q}{t} =  \pdv[2]{Q}{x}+\beta P(x,t)S(x,t) - \beta\gamma Q(x,t)\,,
\end{equation}
which is the same as for the set of PDEs in \cref{subeq:MODEL}.

\section{Front propagation speed for general model \label{app: front speed general model}}
Here we derive the front speed for the general model, \cref{subeq:MODEL_dimensional_system}, using the theoretical framework from ref. \cite{sarloos_frontPropRecent_2003}. Non-dimensionalising, and rescaling as in the main text, we obtain
\begin{subequations}
\label{subeq:WAVE_transofrmed_system_1}
\begin{align}
\pdv{\sigma}{t} &= - p \sigma + \pdv[2]{\sigma}{x}\,,\label{eq:WAVE_transofrmed_system_s}\\
e^{-\mu t}\pdv{p}{t} &= -p (\sigma+\theta)
+ n\int_{-\infty}^t \dd{\tau}\Lambda(t-\tau) \int_\infty^\infty \dd{y} {G}(x-y,t-\tau)e^{-\mu(t- \tau)} p(y,\tau)\sigma(y,\tau)\,,\label{eq:WAVE_transofrmed_system_p}\\
\theta(x,t) &= \int_{-\infty}^t \dd{\tau}  \qty(1-\int_0^{t-\tau}\dd{\tau'}\Lambda(\tau'))\int_\infty^\infty \dd{y} {G}(x-y,t-\tau)e^{-\mu(t- \tau)}
    p(y,\tau) \sigma(y,\tau) \,,
\end{align}
\end{subequations}
where $\sigma=e^{-\mu t}s$, $\theta=e^{-\mu t}q$ and $G(x,t)$ is defined in \cref{eq:infectives_diffusion_kernel}. In the following, the Fourier transform is defined as
\begin{equation}
    \tilde{\phi}(k,t)=\int_{-\infty}^{\infty} \dd{x}e^{-ikx} \phi(x,t)\,,
\end{equation}
and the Laplace-Fourier (LF) transform is defined as
\begin{equation}
    \hat{\phi}(k,\omega)=\int_{0}^{\infty}\dd{t}e^{i\omega t} \tilde{\phi}(k,t)\,.
\end{equation}
This choice of Laplace transform facilitates the identification of wave modes $e^{-i(k x -\omega t)}$. We begin by linearising about a wave propagating into the unstable state, $\{\sigma=1,\theta=0,p=0\}$, to obatin
\begin{subequations}
\label{subeq:WAVE_transofrmed_system_linearised}
\begin{align}
\pdv{\delta \sigma}{t} &= - \sigma_0 \delta p  + \pdv[2]{\sigma}{x}\label{eq:WAVE_solving_system_sigma2_linearised}\,,\\
\delta p(x,t) &= n\int_{0}^t \dd{\tau}\Lambda(t-\tau) \int_\infty^\infty \dd{y} \mathcal{G}(x-y,t-\tau)e^{-\mu(t- \tau)} \delta p(y,\tau)\,,\label{eq:WAVE_solving_system_p2_linearised}
\end{align}
\end{subequations}
which upon LF transforming becomes
\begin{equation}
\hat{p}(k,\omega)=n\hat{\Lambda}(\omega')\hat{p}(k,\omega)\,,
\end{equation}
where $\omega'=\omega + i (k^2 + \mu)$
This has the non-trivial solution, $n\hat{\Lambda}(\omega')=1$, which defines the dispersion relation $\omega=\omega(k)$. Dispersion relations correspond to the poles of the general Green's function which governs system evolution. At large times after the perturbation, the front speed is given by
\begin{subequations}
\label{subeq:WAVE_sarloos_conditions}
\begin{align}
    \Im(\omega - k^* v^*) &= 0\,,\label{eq:WAVE_sarloos_correct_v}\\
    \pdv{\omega}{k}\eval_{k^*} - v^* &=0 \label{eq:WAVE_sarloos_saddle_k}\,.
\end{align}
\end{subequations}
\Cref{eq:WAVE_sarloos_saddle_k} is the saddle point condition for selected wave vector, $k^*$, when solving for the asymptotic form of the Green's function \cite{vansarloos_frontprop_2001}. \Cref{eq:WAVE_sarloos_correct_v} is the condition that the selected front velocity $v^*$ does not change the magnitude of the Green's function to leading order. The conditions in \cref{subeq:WAVE_sarloos_conditions} encode choosing the dominant decay rate, $\lambda=\Im k^*$, at large times and selecting the stationary frame given by $x-v^*t$. 

Because uniform fronts are expected, we must have that $\Re \omega=\Re k=0$, otherwise the front will have an oscillatory envelope, leading to the possibility of negative solutions which are unphysical. Taking $\omega=is,k=i\lambda$ for convenience, we must solve
\begin{equation}
\label{eq:WAVE_laplace_pdf_condition2}
   1=n \hat{\Lambda}(s')\,,
\end{equation}
where $s'=s-\lambda^2+\mu$. The conditions \cref{subeq:WAVE_sarloos_conditions} to determine $v^*$ become
\begin{align}
\label{eq:WAVE_VS_conditions_for_delay_process_12}
    s = v^* \lambda \qq{and} \pdv{s}{\lambda} = v^*\,,
\end{align}
which leads to
\begin{align}
    1 = n \hat{\Lambda}(s')\,, \qq{and}
    0 = (v^*-2\lambda)\dv{\hat{\Lambda}}{s'}
    \label{subeq:WAVE_cum_prob_cond_22}\,,
\end{align}
with $s'= v^* \lambda -\lambda^2 +\mu$. As the function $\hat{\Lambda}(s)$ is monotonically decreasing, there is one unique solution to the system \cref{subeq:WAVE_cum_prob_cond_22} at $v^*=2\lambda$. Call this solution $s=s'_*$ and so
\begin{equation}
    s'_* = \frac{{v^*}^2}{4} +\mu\,.
\end{equation}
The general front velocity for this scenario has the form
\begin{equation}
\label{eq:WAVE_gen_speed}
    v^* = 2\sqrt{s'_*(n)-\mu}\,,
\end{equation}
which is the selected front velocity. The dependence of $s'_*$ on $n$ is included as altering the burst size will directly affect the front speed by changing the available solutions. To take some common examples of lysis distributions, a delta distribution, $\Lambda(t)=\delta(t-\tau)$, gives the front propagation speed as
\begin{equation}
\label{eq:APP_delta_dist}
    v=2\sqrt{\frac{\ln(n)}{\tau} - \mu}\,,
\end{equation}
an exponential distribution, \cref{eq:app_exp_lysis}, gives,
\begin{equation}
\label{eq:APP_exp_dist}
    v=2\sqrt{\gamma(n-1) - \mu}\,,
\end{equation}
which agrees with \cref{eq:RandD_scaled_velocity}, whereas a Gamma distribution,
\begin{equation}
    f_{\Gamma}(x;\alpha,\beta) = \frac{\beta t^{\alpha -1} e^{-\beta t}}{\Gamma(\alpha)}\,,
\end{equation}
gives
\begin{equation}
\label{eq:APP_gamma_dist}
    v=2\sqrt{\beta\qty({n}^{1/\alpha}-1)-\mu}\,.
\end{equation}
Taking the limit $\beta\rightarrow \infty$ in such a way that $\flatfrac{\alpha}{\beta}=\tau=\text{const}$ reduces the Gamma distribution to $\delta(t-\tau)$ which, applying to \cref{eq:APP_gamma_dist}, gives \cref{eq:APP_delta_dist}. The delta distribution may be shown to exhibit the minimum speed for this system for any viable probability distribution (i.e. probability distributions defined on the positive reals) centred on the same mean $\tau$ \cite{GAVAGNIN201991}, which makes it an interesting case study.

\section{Infection Model without Bacterial Growth}\label{sec:APP_no_growth_wave_speed}
Here, we solve the model from the main text without growth ($\mu=0$) but in the limit of very large initial concentration $S_0$ to derive the non-growing wavespeed $c_0$. Eq. (2) from the main text with $\mu=0$ is
\begin{subequations}
	\label{subeq:MODEL_nondim_no_growth}
	\begin{align}
	\underbrace{\pdv{s}{t}}_{\mathrm{Susceptibles}} &= - {\underbrace{ps}_{\mathrm {Infection}}} + {\underbrace{\pdv[2]{s}{x}}_{\mathrm{Swimming}}}\,,\\
	\underbrace{\pdv{q}{t}}_{\mathrm{Infected}} &= {\underbrace{ps}_{\mathrm{Infection}}} - {\underbrace{\gamma q}_{\mathrm{Lysis}}} + {\underbrace{\pdv[2]{q}{x}}_{\mathrm{Swimming}}}\,,\label{subeq:no_grow_q}\\
	\underbrace{\pdv{p}{t}}_{\mathrm{Phage}} &= {\underbrace{n \gamma q}_{\mathrm{Lysis}}}- {\underbrace{p s}_{\mathrm{Infection}}}  - {\underbrace{p q}_{\mathrm{Super\text{-}infection}}} \,.
	\end{align}
\end{subequations}
Taking the limit of a large susceptible population, the term $ps$ dominates and we find there is a separation of timescales which causes the phage to rapidly converge to a dynamic equilibrium i.e. $\pdv{p}{t}=0$ and so
\begin{equation}
p=\frac{n\gamma q}{s+q}\,,
\end{equation}
which is eq. (3) from the main text. Substituting this into \cref{subeq:no_grow_q} gives
\begin{subequations}
\label{subeq:no_grow_linearised}
\begin{align}
\pdv{s}{t} &= -\frac{n\gamma q s}{s+q} + \pdv[2]{s}{x}\,,\\
\pdv{q}{t} &= \frac{n\gamma q s}{s+q}-\gamma q + \pdv[2]{q}{x}\,,
\end{align}
\end{subequations}
which has the form of eq. (5) in the main text with $\sigma\rightarrow s,~\theta\rightarrow q,~\mu_s=0$, and hence the wave speed $c_0=\sqrt{(n-1)\gamma}$ follows by using standard techniques~\cite{sarloos_frontPropRecent_2003} as in the main text. We see that $c_0=\lim_{\mu\rightarrow 0}c_-(\mu)$, but we emphasise that this only holds in the limit of large susceptible populations that we take here. An alternative derivation is to find all possible front speeds of \cref{subeq:MODEL_nondim_no_growth}, where in the limit $S_0\rightarrow\infty$, the only physically possible speed is $c_0$.

\section{Simulation Details}\label{sec:APP_SIM_DETAILS}

We split the following into a discussion of the simulation implementations, followed by a discussion of the speed calculation. The simulation implementations are divided into the simple model simulations in \cref{sec:APP_simple_model_sim_details}, and the more involved simulation details for the realistic model in \cref{sec:APP_realistic_model_sim_details}. In \cref{sec:APP_sim_details_speed_calc}, we describe our speed calculation method. We were required to use this atypical method due to numerical issues with constant height tracking. All simulations and calculations were performed on a Dell XPS 15 7590 with i7-9750H CPU and 16Gb RAM.

\subsection{Simple model simulation details}\label{sec:APP_simple_model_sim_details}
\Cref{subeq:MODEL_nondim} was simulated in Python using the
\texttt{solve\_ivp} module from SciPy \cite{2020SciPy-NMeth}. The implicit
backward differentiation formula method was used with a time step adaptively chosen to satisfy absolute and relative error tolerances of $10^{-8}$ and $10^{-5}$, respectively \cite{shampine_matlab_1997}. Using typical experimental parameters gives a domain length $l_{sim} \sim$ \SIrange{4}{10}{\cm}. In non-dimensional parameters, the domain length was $\order{100}$. This domain length was chosen as a compromise between being long enough such that the wave fronts could fully develop whilst not being excessively numerically demanding. Space was uniformly discretised into $\order{10^4}$ collocation points.

The initial conditions were $[s,q]=[1,0]$ everywhere, with a smooth step-like bacteriophage profile near the origin $p(t=0)=p_0/(1+e^{\Upsilon(x-x_0)})$, with amplitude $p_0=5$, steepness $\Upsilon=10$, and width $x_0=l_{sim}/5$. Zero-flux boundary conditions apply at the domain edges. Typical simulations for $\mu \leq 1$ took \SIrange{30}{60}{\second}.

\subsection{Realistic model simulation details}\label{sec:APP_realistic_model_sim_details}
In this section, we describe the numerical solution of the realistic system, which incorporates two additional infected stages after the delay (eclipse period), of length $T_e$. These new stages of infection decay exponentially with the same decay rate, $r$. In total, this has the effect of approximating the lysis distribution as a shifted Gamma distribution $\Lambda(T)\sim\Gamma(T,2;r,T_e)$ (\cref{eq:shiftedGamma} with $\alpha=2$). In the following, we use the non-dimensional equivalent, $\Lambda(t)\sim\Gamma(t;2,\gamma,\tau_e)$ with $\tau_e=\beta S_0T_e$ and $\gamma=r/(\beta S_0)$. The total number of infected bacteria is the sum of these infected stages. The system could be written as
\begin{subequations}
\label{subeq:modified_PDDE_system_full}
\begin{align}
\pdv{\sigma}{t} &= -p\sigma+\pdv[2]{\sigma}{x}\,,\\
\pdv{\theta_0}{t} &= \pdv[2]{\theta_0}{x}+p\sigma-\mu \theta_0 + \pdv[2]{\theta_1}{x}-\int_{0}^t \dd{\tau}\delta(t-\tau-\tau_e) e^{\mu \qty(\tau-t)}\int_\infty^\infty \dd{y} G(x-y,t-\tau) p(\tau,y)\sigma(\tau,y)\,,\\\\
\pdv{\theta_1}{t} &= -\gamma \theta_1 -\mu \theta_1 + \pdv[2]{\theta_1}{x}+\int_{0}^t \dd{\tau}\delta(t-\tau-\tau_e) e^{\mu \qty(\tau-t)}\int_\infty^\infty \dd{y} G(x-y,t-\tau) p(\tau,y)\sigma(\tau,y)\,,\\
\pdv{\theta_2}{t} &= \gamma (\theta_{1} - \theta_2) -\mu \theta_2 + \pdv[2]{\theta_2}{x}\,,\\
\pdv{p}{t} &= e^{\mu t} \qty(n\gamma \theta_2 - p \qty(\sigma+\theta_1+\theta_2+\theta_0))+\delta_p \pdv[2]{p}{x}\,,
\end{align}
\end{subequations}
where the total number of infected bacteria is $\theta=\theta_0+\theta_1+\theta_2$. This is a simplification of the system in \cref{model} but also includes phage diffusion with diffusion coefficient $\delta_p\ll 1$. This is included for more generality and to better approximate reality, although for all practical purposes we do not expect this to have a significant effect on the system dynamics. 
We now split the solution method into blocks of time. In the following, the age of infectives is $a$ and current time is $t$. In the first block we have $a<t$ (age of infected is always less than the current time) which corresponds to  $0\leq t \leq \tau_e$. Hence, we solve
\begin{subequations}
\begin{align}
\pdv{\sigma}{t} &= -p\sigma+\pdv[2]{\sigma}{x}\,,\\
\pdv{\theta_0}{t} &=  p\sigma-\mu \theta_0 + \pdv[2]{\theta_0}{x}\,,\\
\pdv{\theta_1}{t} &= -\gamma \theta_1 -\mu \theta_1 + \pdv[2]{\theta_1}{x}\,,\\
\pdv{\theta_2}{t} &= \gamma (\theta_{1} - \theta_2) -\mu \theta_2 + \pdv[2]{\theta_2}{x}\,,\\
\pdv{p}{t} &= e^{\mu t} \qty(n\gamma \theta_2 - p \qty(\sigma+\theta_1+\theta_2))+\delta_p \pdv[2]{p}{x}\,,
\end{align}
\end{subequations}
and in the successive time blocks, $j\tau_e \leq t \leq (j+1)\tau_e$ for integers $j\geq 1$, we have
\begin{subequations}
\begin{align}
	\pdv{\sigma}{t} &= -p\sigma+\pdv[2]{\sigma}{x}\,,\\
	\pdv{\theta_0}{t} &=  p\sigma-\mu \theta_0 - N(x,t) + \pdv[2]{\theta_1}{x}\,,\\
	\pdv{\theta_1}{t} &= -\gamma \theta_1 -\mu \theta_1 + N(x,t) + \pdv[2]{\theta_1}{x}\,,\\
	\pdv{\theta_2}{t} &= \gamma (\theta_{1} - \theta_2) -\mu \theta_2 + \pdv[2]{\theta_2}{x}\,,\\
	\pdv{p}{t} &= e^{\mu t} \qty(n\gamma \theta_2 - p \qty(\sigma+\theta_0+\theta_1+\theta_2))+\delta_p \pdv[2]{p}{x}\,,\\
    N(x,t) \equiv u(x,\tau_e) &= e^{-\mu \tau_e}\int_0^L\dd{y} G(x-y,\tau_e)p(y,t-\tau_e)\sigma(y,t-\tau_e)\,,\\
    \pdv{u}{s} &= \pdv[2]{u}{x} - \mu u \qq{with} u(0,x)=p(t-\tau_e,x)\sigma(t-\tau_e,x)\,. \label{eq:NA_JonesPropU}
\end{align}
\end{subequations}

\subsection{Front propagation speed calculation}\label{sec:APP_sim_details_speed_calc}
To obtain numerical wave speeds, the typical approach is to track the dynamics $x(t)$ of the point $x$ where the given population ($\phi=q,~p,~\theta$ etc.) crosses some constant amplitude $h$. We choose an amplitude $h\ll 1$, so that the points tracked are all located at the front edge of the profile \cite{sarloos_frontPropRecent_2003} (for the susceptibles, $\phi=e^{\mu_s t}-s$ and $\phi=1-\sigma$ are used so that $\lim_{x\rightarrow\infty}\phi=0$). In discrete terms, the wave dynamics $x_i=x(t_i)$ is the solution of $\phi(x_i,t_i)=h$ for some series of output times $t_i$, with $i$ specifying the discrete time index. The calculated front speed is then
\begin{equation}
c_i(t_{i+1})=\frac{x_{i+1}-x_{i}}{t_{i+1}-t_{i}}\,,\\
\label{eq:numerical_front_speed}
\end{equation}
for this period. If $t_{i+1}-t_{i}\ll 1$, this approaches the instantaneous front speed.

\begin{figure}[h!]
\includegraphics[width=0.5\linewidth]{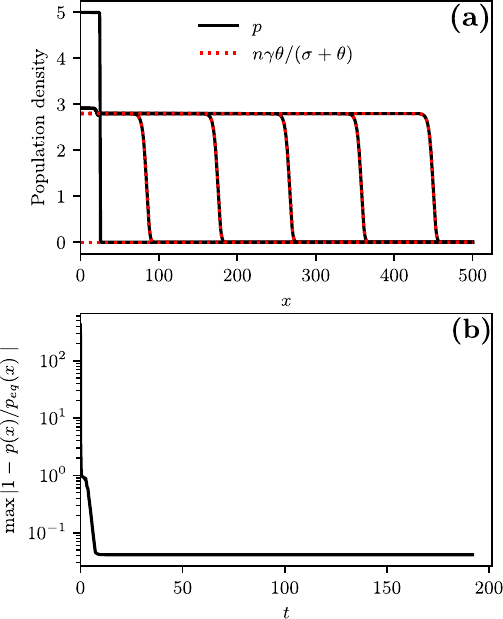}
\caption{(a) Numerical solution to eq. (2) in the main text for the default parameter set. The dynamic equilibrium of the phage, eq. (3) from the main text is also plotted from the numerical solutions for $\sigma,~\theta$. The convergence to this dynamic equilibrum is seen to be excellent for the majority of the domain, with a small exception near the origin. The parameters were $n=4,\gamma=0.7,\mu_s=0.8$. (b) The maximum fractional discrepancy, $|1-p/p_{eq}|$ over the whole domain for each timestep was calculated for the data from (a). This was used as an indication of how well the dynamic equilibrium matched full phage profile. Despite the discrepancy at the origin, the agreement is seen to be excellent.  \label{fig:phg_eqm}}
\end{figure}

This approach was successful for the scaled populations $\phi=\theta,~1-\sigma,~p$ but failed for the un-scaled populations $\phi=q,~e^{\mu_s t}-s$. This is because these are exponentially increasing populations and the population level $h$ becomes similar to the numerical error in these populations at long times. Hence, for these populations we employed the following iterative procedure.  For any two successive times, $t_i$ and $t_{i+1}$
\begin{enumerate}
\item Choose a small relative height
\begin{equation*}
	h_i=\varepsilon\frac{ \max_x\{ \phi(t_{i+1},x) \} + \max_x\{ \phi(t_{i},x) \} }{2}\,,
\end{equation*}
where $\varepsilon\ll1$ (we typically use $\varepsilon = 10^{-8}$) and $ \max_x\{ \phi(t_{i},x) \}$ is the maximum value of the profile $\phi(x,t)$ over $x$ at the time $t=t_i$, i.e. the height $h_i(t_i)$ is adjusted to be a fixed small fraction of the average of the maxima of the profiles at $t_i$ and $t_{i+1}$.
\item Solve $\phi(x_i,t_i)=h_i$, $\phi(x_{i+1},t_{i+1})=h_i$ for $x_{i}$, $x_{i+1}$.
\item Calculate the front speed $c_i$ in the period $t\in (t_i,t_{i+1})$ from \cref{eq:numerical_front_speed}.
\end{enumerate}

\section{Rapid Relaxation towards the Steady State of the Phage Population}\label{sec:APP_phg_eqm}
In the main text we argue that the phage population will reach dynamic equilibrium much faster than the bacterial populations. This yields eq. (3) in the main text, which is
\begin{equation}
p=\frac{n\gamma q}{s+q}=\frac{n\gamma \theta}{\sigma+\theta}\,,
\label{eq:phg_eqm}
\end{equation}
where $\sigma=e^{-\mu_s}s,~\theta=e^{-\mu_s}q$. In \cref{fig:phg_eqm}(a) we show the comparison between numerical solutions for $p$ and the dynamic equilibrium \cref{eq:phg_eqm} for a number of output times and in \cref{fig:phg_eqm}(b) we evaluate the maximum fractional difference for all times simulated. 

\section{Absence of superinfection}\label{sec:APP_no_superinfection}
\begin{figure}[h!]
\includegraphics[width=0.5\linewidth]{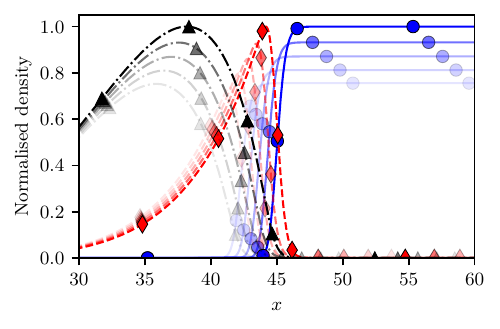}
\caption{Numerical solution of \cref{subeq:MODEL_nondim} without superinfection. The population profiles have been normalised by their respective maximum of the last outputted timestep in the plot. To show the time evolution, earlier times have greater transparency.  \label{fig:no_superinfection}}
\end{figure}

Here, we present the effect on the simple model \cref{subeq:MODEL_nondim} of removing the superinfection term in \cref{fig:no_superinfection}. Under such conditions, the phage grow unboundedly. As this only affects the bulk of the wave behind the front, there is no effect on the front speed. However, the phage wave as a whole is no longer self-similar. 

\section{Absence of Wave Propagation in Regime Dominated by Exponential Growth of Bacteria}\label{sec:APP_mu_greater_1}
\begin{figure}[h!]
\includegraphics[width=0.5\linewidth]{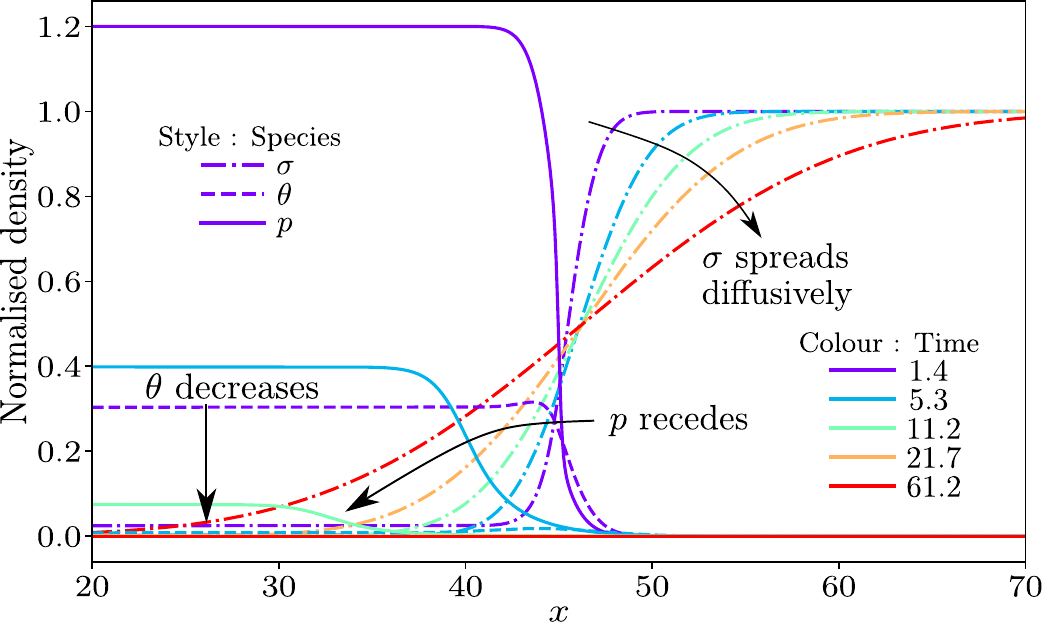}
\caption{For growth rates which are too high (such that $\mu>1$), travelling fronts do not form and the profiles spread out whilst decreasing, in a manner qualitatively similar to diffusion. The effects are easiest to see in the rescaled system. Arrows are included above to guide the eye in the evolution of the different populations. The susceptible population, $\sigma$ is seen to invade the phage, $p$ as the virus replicates slower than the rate at which new susceptible bacteria are produced, causing the virus population to asymptotically tend to 0. The infected bacteria, $\theta$, follow the same pattern. The spreading dynamics of all populations are diffusion-like and no travelling fronts form. The parameters here were $\mu_s=1, \gamma=0.01, n=50$ which gives $\mu\approx2$. \label{fig:diff_branch}}
\end{figure}

For $\mu>1$ one obtains a characteristically different behaviour as we cannot solve for uniformly translating front solutions. \Cref{fig:diff_branch} shows a numerical solution under these conditions, where there is clearly no front formation. The profile is qualitatively diffusive as the $\theta$ and $p$ species cannot reproduce quickly enough to catch up with the population expansion of the susceptibles; hence the evolution of the system is dominated by the diffusive dynamics of $\sigma$. The number of infected bacteria, $\theta$, is also seen to decrease rapidly.

\section{Infection Model Including Phage Diffusion and Death}\label{sec:APP_extended_simple_model}
\begin{figure}[h!]
	\includegraphics[width=\linewidth]{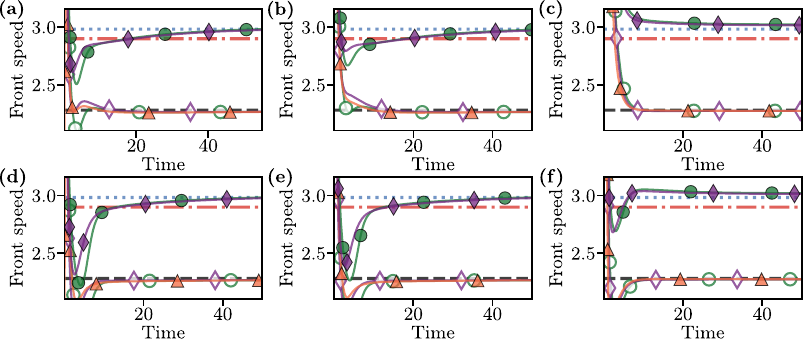}
\caption{Convergence of the front speed to the theoretical predictions tested for two orders of magnitude of phage diffusion. Although there are qualitative differences in the convergence properties the long-time limit of the front speed is independent of phage diffusion.  In all cases the front speed was evaluated with $\varepsilon=10^{-5}$, and with the exception of $\delta_p$, all the parameters are the default set, $\delta_s=1$, $\delta_q=1$, $n=4$, $\mu=0.8$, $\gamma=0.7$, $\nu=0$, $\varepsilon=10^{-5}$ and $\Delta x = 1/100$.
(a)\label{fig:PHAGE_DIFF_DEATH} Phage with slightly higher than realistic diffusion coefficient of \SI{6.5}{\micro\meter^2\second^{-1}}, giving $\delta_p = 5\times10^{-2}$.
	(b) Phage with unrealistically high diffusion coefficient, only half that of the bacteria with \SI{65}{\micro\meter^2\second^{-1}}, giving $\delta_p = 0.5$.
(c) Exceptionally high rate of phage diffusion, \SI{650}{\micro\meter^2\second^{-1}}, giving $\delta_p = 5$. (d) is as in (a) but has $\nu=5$ ($m=$\SI{12.5}{\hour^{-1}} in dimensional units) which is an unrealistically high rate of phage death. (e) and (f) are then as (b) and (c) but also have $\nu=5$. As before, the inclusion of a very high phage death rate, even in combination with large phage diffusion only affects the convergence to the expected speeds.}
\end{figure}

In this section, the system investigated in the main text is extended to allow for different diffusivities of the susceptible and infected classes; a non-zero phage diffusivity; and a spontaneous phage death term. We find the front-speed predictions from the main text are still realised. The extended non-dimensional model is
\begin{subequations}
\label{subeq:APP_MODEL_nondim}
\begin{gather}
\underbrace{\pdv{s}{t}}_{\mathrm{Susceptible}} = {\underbrace{\mu_s s}_{\mathrm{Growth}}} - {\underbrace{ps}_{\mathrm{Infection}}} + {\underbrace{\delta_s\pdv{^2s}{x^2}}_{\mathrm{Swimming}}}\,,\\
{
\underbrace{\pdv{q}{t}}_{\mathrm{Infected}} = {\underbrace{ps}_{\mathrm{Infection}}} - {\underbrace{\gamma q}_{\mathrm{Lysis}}} + {\underbrace{\pdv{^2q}{x^2}}_{\mathrm{Swimming}}}\label{subeq:APP_MODEL_infected}}\,,\\
\underbrace{\pdv{p}{t}}_{\mathrm{Phage}} = {\underbrace{n \gamma q}_{\mathrm{Lysis}}} - {\underbrace{p s}_{\mathrm{Infection}}} - {\underbrace{p q}_{\mathrm{Super\text{-}infection}}}-\underbrace{\nu p}_{\mathrm{Death}} + {\underbrace{\delta_p\pdv{^2q}{x^2}}_{\mathrm{Diffusion}}}\,,
\end{gather}
\end{subequations}
where $\nu=m/(\beta S_0)$ with $m$ the dimensional phage death rate, $\delta_s=D_s/D_q$ and $\delta_p=D_p/D_q$. We re-scale the system as before using $\sigma=se^{-\mu_s t}$ and $\theta=qe^{-\mu_s t}$ to obtain
\begin{align}
\label{eq: p approx 2}
\pdv{p}{t}=e^{\mu_s t}\qty[n\gamma\theta - p(\sigma+\theta)] - \nu p+ \delta_p \pdv[2]{p}{x} \,.
\end{align}
where we see that the death and diffusion terms of the phage are much smaller than the infection and lysis terms as $t\rightarrow\infty$. We can again approximate $p$ by its steady-state value
\begin{align}
\label{eq: p approx}
p\sim n\gamma q/(s+q) = n\gamma \theta/(\sigma+\theta)\,.
\end{align}
Inserting this expression into \cref{subeq:APP_MODEL_infected} after appropriate re-scaling gives
\begin{subequations}
\label{subeq:APP_RandD_scaled_system2}
\begin{align}
\pdv{\sigma}{t} &= -\frac{n\gamma\theta\sigma}{\theta+\sigma} + \delta_s\pdv[2]{\sigma}{x}\,,\\
\pdv{\theta}{t} &= \frac{n\gamma\theta\sigma}{\theta+\sigma}-\qty(\gamma+\mu_s) \theta + \pdv[2]{\theta}{x}\,,
\end{align}
\end{subequations}
again giving two coupled FKPP equations, which is identical to eq. (5) in the main text and therefore yields the same speed $c_-$, with $c_+$ following similarly. We also numerically evaluated the speed with a range of phage diffusivities up to $\delta_p=5$ and a phage death rate of $\nu=5$  (\cref{fig:PHAGE_DIFF_DEATH}). In all cases, the long-time numerical speed matches the predicted speed, with no impact of the phage diffusivity or death rate.

\section{Delay-differential system with no phage diffusion}\label{sec:APP_PIDE_no_phage_diffusion}
\begin{figure}[h!]
\includegraphics[width=\linewidth]{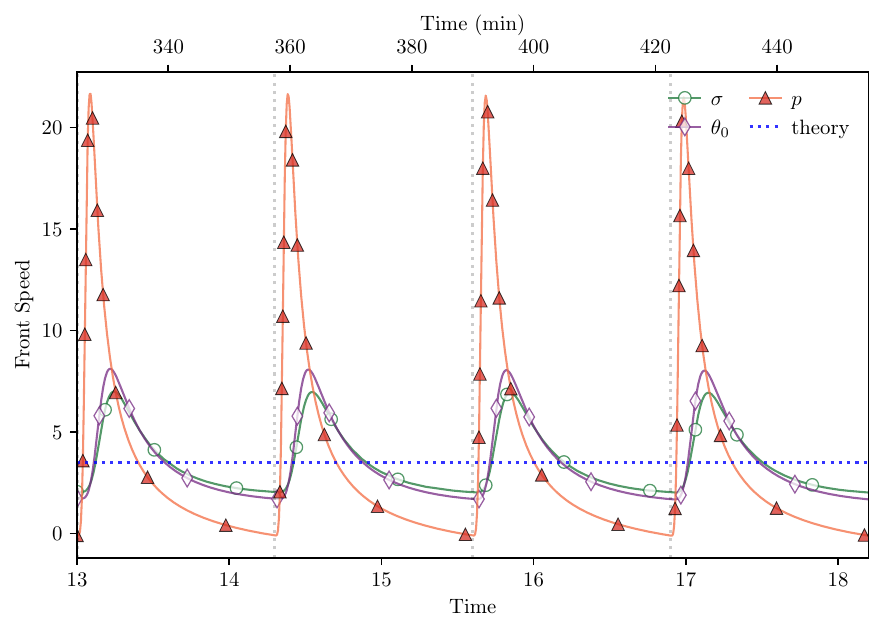}
\caption{
Pure delay differential equations with lysis time $L=\SI{32.5}{\minute}$ and burst size $n=150$, with remaining parameters those the defaults from \cref{tbl:parameters}. Large oscillations in the wave speed are sustained for as long as we simulated. The average simulation speed matches the theoretical prediction. \label{fig:APP_PIDE_no_phage_diffusion}
}
\end{figure}

For a pure delay-differential equation solution to \cref{subeq:MODEL_dimensional_system}, discussed in \cref{realistic}, we find sustained oscillations which visually do not appear to decrease in amplitude over time, see \cref{fig:APP_PIDE_no_phage_diffusion}. In spite of this, the smoothed speed, extracted using a sliding average over a full lysis period still gives the expected theoretical speed.

\section{Wave speed}\label{sec:APP_PIDE_full_wave_speeds}
\begin{figure}[h!]
\includegraphics[width=0.5\linewidth]{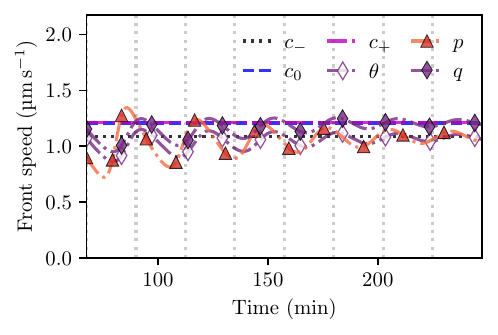}
\caption{Numerically calculated front speeds for \cref{subeq:MODEL_dimensional_system}, showing the convergence to the theoretically predicted speeds without smoothing. Despite sustained, yet decreasing oscillations, the front speeds are seen to converge to the theoretically predicted values. Simulation performed with parameters from \cref{tbl:parameters}.  \label{fig:APP_PIDE_full_wave_speeds}
}
\end{figure}

In the main text, we plotted the smoothed front speeds in \cref{fig:DDE_self_similar_speeds} for the realistic model \cref{subeq:MODEL_dimensional_system} using the default parameters from \cref{tbl:parameters}. In \cref{fig:APP_PIDE_full_wave_speeds}, we plot the original front speeds to demonstrate the oscillations about the convergence to the theoretical speeds. The oscillations make it difficult to see the two separate speeds the waves are converging to. As the amplitude of the oscillations decrease with time, the speed the populations are converging to becomes more obvious at the end of the plotted period.

\cleardoublepage
\bibliographystyle{apsrev4-2}
\bibliography{ms}

\end{document}